\documentclass[reprint,aip]{revtex4-1}
\usepackage{graphicx}
\usepackage{dcolumn}
\usepackage{bm}
\usepackage{amsfonts,amssymb,amsmath,geometry, url,siunitx}
\usepackage{gensymb}
\usepackage{graphicx}
\usepackage{hyperref}
\usepackage{setspace}
\usepackage{hyperref}
\hypersetup{
  colorlinks=true,    
   linkcolor=blue,       
   citecolor=blue,       
   filecolor=black,        
   urlcolor=blue,        
   linktoc=page            
}

\begin{document}
\title{Spectroscopic signatures of native charge compensation in Mg doped GaN Nanorods}
\author{Rajendra Kumar}
\thanks{These two authors contributed equally}
\affiliation{Chemistry and Physics of Materials Unit\\Jawaharlal Nehru Centre for Advanced Scientific Research (JNCASR), Bangalore-560064, India}
\author{Sanjay  Nayak}
\thanks{These two authors contributed equally}
\affiliation{Chemistry and Physics of Materials Unit\\Jawaharlal Nehru Centre for Advanced Scientific Research (JNCASR), Bangalore-560064, India}
\author{S.M. Shivaprasad}
\email{smsprasad@jncasr.ac.in}
\affiliation{Chemistry and Physics of Materials Unit\\Jawaharlal Nehru Centre for Advanced Scientific Research (JNCASR), Bangalore-560064, India}


\begin {abstract}
We study the native charge compensation effect in Mg doped GaN nanorods (NRs), grown by Plasma Assisted Molecular Beam Epitaxy (PAMBE), using Raman, photoluminescence (PL) and X-ray photoelectron spectroscopies (XPS). The XPS valence band analysis shows that upon Mg incorporation the E$_F$-E$_{VBM}$ reduces, suggesting the compensation of the native n-type character of GaN NRs. Raman spectroscopic studies on these samples reveal that the line shape of longitudinal phonon plasmon (LPP) coupled mode is sensitive to Mg concentration and hence to background n-type carrier density. We estimate a two order of native charge compensation in GaN NRs upon Mg-doping with a concentration of 10$^{19}$-10$^{20}$ atoms cm$^{-3}$. Room temperature (RT) PL measurements and our previous electronic structure calculations  are used to identify the atomistic origin of this compensation effect.

\end {abstract}


\maketitle

Epitaxially grown semiconductor nanostructures such as quantum dots, nanowires and nanorods of III-V materials have been investigated extensively in the literature\cite{ihn2006gaas,li2008growth,cerutti2006wurtzite,maartensson2004epitaxial}. The interest of studies on these nanostructured materials lies in the facts that it offer an extra degree of freedom to manipulate it's material properties in comparison to bulk form. The enhanced performance of nanostructure based devices is partly due to the effective  lateral  stress  relaxation as a consequence of the  presence  of  facet  edges  and  sidewalls. These crystal facets also minimize or eliminate the formation of dislocations, and consequently leads to the fabrication of nearly defect-free III-V  semiconductor  nanostructures with large surface to volume ratios\cite{glas2006critical,verheijen2006growth}. 1D NRs offer several advantages over the planar structure such as reduced dislocation density\cite{lin2010ingan,bengoechea2014light}, lower polarization field\cite{wang2008gan}, and enhanced light output efficiency\cite{nguyen2011p}. This approach can also reduce the cost of LED fabrication on large-area Si substrates. Thus, Growth, characterization and optimization of single crystalline nitride NRs  have been of great interest, which has the tremendous potential for technological applications. 
\par
Various techniques such as Hall measurements, C-V measurements etc, have been employed to study the electronic properties like carrier density, mobility etc., of semiconducting films. These techniques require preparation of Ohmic contacts on the films, which is very difficult for nanostructures such as NRs, NWs etc., due to their discontinuity and small dimensions. Raman spectroscopy is a very powerful technique to study material properties of nitride semiconductors. It is well known that formed GaN films are intrinsically n-type and the background carrier densities are to the order of 10$^{17}$-10$^{19}$ cm$^{-3}$. Often the oscillation of such high free electron densities  are collective and referred as \textit{plasmons}. The plasmon oscillations of these free carriers couple with Raman active longitudinal optical (LO) phonon modes via its associated longitudinal electric field gives rise to longitudinal phonon-plasmon(LPP) coupled mode, which is Raman active. Behaviour of this mode drastically changes with carrier concentration, enabling a contactless, local, optical probe of carrier concentration. The shift of the peak position of the LPP mode upon Mg doping have been reported earlier\cite{wang2014p}, but the effect on line shape is not clearly understood.
\par
In this work we synthesize single crystalline (wurtzite)  hexagonal shaped GaN NRs on Si (111) surface using plasma assisted molecular beam epitaxy (PA-MBE). The Mg and Ga flux rates are varied by adjusting respective K-cell temperatures and monitored by measuring from the beam equivalent pressure (BEP). Concentrations of Mg incorporated in these samples are estimated by SIMS measurements as reported in our previous work\cite{nayak2018enhanced}. Other growth parameters employed can be found elsewhere\cite{nayak2018enhanced}. The morphology of the grown films are determined \textit{ex-situ} using a field emission scanning electron microscope (FESEM, Quanta 3D operating at 20 kV). The optical properties of the films at RT are studied by photo-luminescence spectroscopy (PL, Horiba Jobin Yvon) using a Xenon lamp source for excitation and Raman spectroscopy is performed with an Ar laser of wavelength 514 nm in the back scattering geometry $\mathrm{z(y,-)\overline{z}}$. The electronic structure of the films are characterized by X-ray photoelectron spectroscopy (XPS)  with Al-K$_\alpha$ (1486.7 eV) source.

\begin{figure}
\centering
\includegraphics[width=0.5\textwidth]{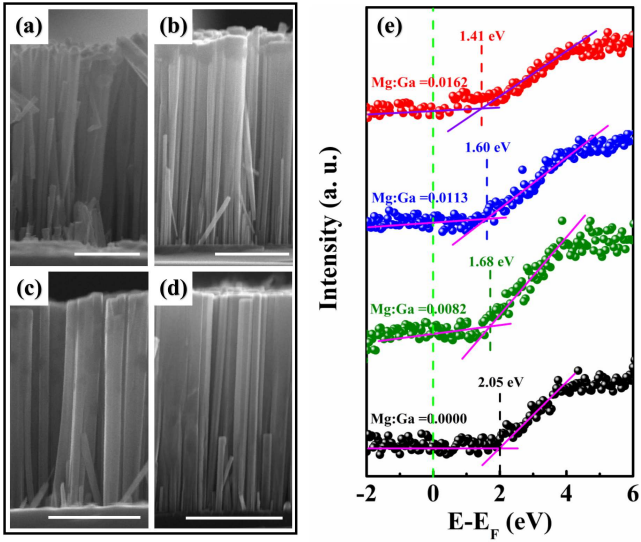}
\caption{(a), (b), (c) and (d) are cross section FESEM images of samples A, B, C and D,respectively. (e) shows valence band spectra of samples A, B, C and D.}
\label{sem_xps}
\end{figure}
\par
The cross section FESEM images of the grown samples as shown in Fig.\ref{sem_xps},  reveal the formation of well aligned uniform NRs with high quality single crystallinity\cite{nayak2018enhanced}. A thorough discussion on their structural and morphological analysis and Mg dopant concentration of these samples has been reported earlier\cite{nayak2018enhanced}. The effect of Mg doping on electronic structure  is characterized by XPS valence band spectra and is shown in Fig.\ref{sem_xps}(e). It is well known in the literature that position of Fermi level ($\mathrm{E_F}$) with respect to Valence Band Maximum ($\mathrm{E_{VBM}}$) is the signature of the type and concentration of the charge carrier. We find that for sample A (undoped),  $\mathrm{E_F}$ is at 2.05 eV above $\mathrm{E_{VBM}}$ indicating native n-type character of pristine GaN NRs as expected in GaN growth. Upon Mg incorporation in the NRs the Fermi level moves towards VBM and the $\mathrm{|E_F - E_{VBM}|}$ reduces to 1.68, 1.60 and 1.41 eV  for samples B, C and D, respectively. This clearly shows that the native n-type character is being compensated upon Mg doping.
\par
\begin{figure}
\centering
 \includegraphics[width=0.45\textwidth]{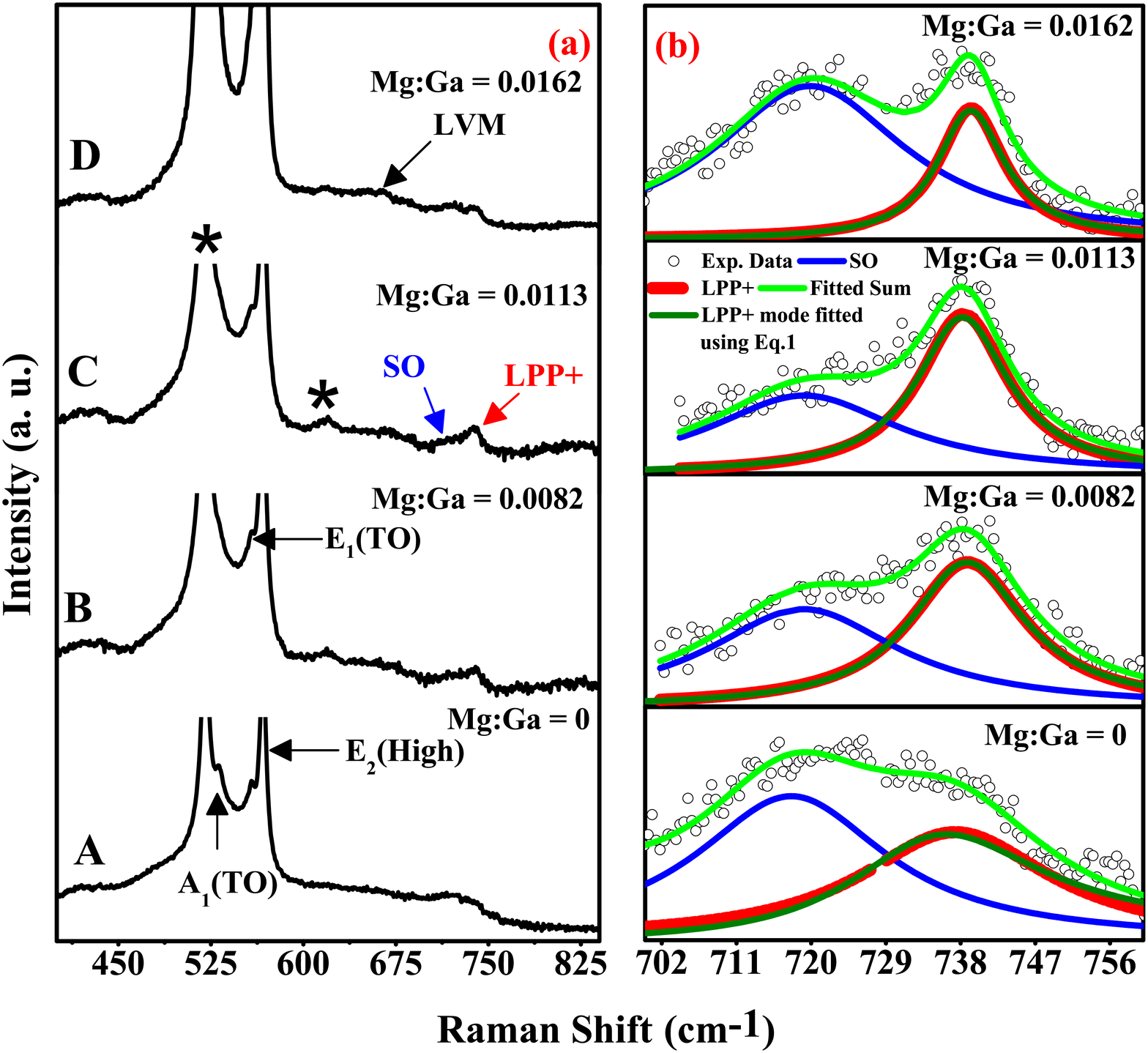}
 \caption{(a) Raman spectra of samples A, B, C and D. (b) shows  the deconvoluted SO and LPP Raman modes of the samples. Experimental data are deconvoluted with Lorentzian functions.}
 \label{LPP}
 \end{figure}

We acquired Raman spectra of all the four samples at RT and are shown in  Fig.~\ref{LPP}.  Presence of 
E$_2$(high) and E$_2$(low) at $\mathrm {\approx 567 \ cm^{-1}}$ and $\mathrm {\approx 144 \ cm^{-1}}$ (not shown here), E$_1$(TO) and A$_1$(TO) at $\mathrm {\approx 557 \ cm^{-1}}$ and  at $\mathrm {\approx 532 \ cm^{-1}}$, respectively, confirms single crystalline wurtzite phase of the GaN films. Along with the expected peaks from GaN, a local vibrational mode (LVM)  appears at 662 cm$^{-1}$ for Mg doped samples, which is attributed to the Mg-N bond\cite{Kaschner1999}. For single crystalline and relaxed thin films, the A$_1$(LO) and E$_1$ (LO) phonon frequencies are reported to be 734 $\mathrm{cm^{-1}}$ and 741  $\mathrm{cm^{-1}}$ , respectively \cite{davydov1998phonon} with a deviation of 1-2 $\mathrm{cm^{-1}}$\cite{azuhata1995polarized,siegle1997zone}. Since in the scattering geometry $\mathrm{z(y,-)\overline{z}}$ only  A$_1$(LO) is allowed, the two different modes observed in our studies  at $\mathrm {\approx 721 \ cm^{-1}}$ and $\mathrm {\approx 738 \ cm^{-1}}$ are identified as  surface optical (SO) phonon peak and  longitudinal phonon-plasmon (LPP+) mode, respectively\cite{robins2016raman}. The SO phonon peak is generally absent in bulk GaN films but is quiet prominent in the NRs due to the relatively large surface to volume ratio. Raman line profile of the LPP coupled mode is given by following equations\cite{cheng2009raman}
\begin{equation}
\mathrm{I(\omega)=const.A(\omega).Im[-\epsilon(\omega)^{-1}]}
\label{I}
\end{equation}
where $\omega$ is Raman shift, $\epsilon(\omega)$ is dielectric function and A($\omega$) is of the following form
\begin{equation}
\begin{split}
A(\omega)=& 1+2C\frac{\omega_{TO}^2}{\delta}[\omega_p^2\gamma(\omega_{TO}^2-\omega^2)-\omega^2\Gamma(\omega^2+\gamma^2-\omega_p^2)] \\
& +C^2\{\omega_p^2[\gamma(\omega_{LO}^2-\omega_{TO}^2)+\Gamma(\omega_p^2-2\omega^2)] \\
& +\omega^2\Gamma(\omega^2+\gamma^2)\}\left[\frac{\omega_{TO}^4}{\delta(\omega_{LO}^2-\omega_{TO}^2)}\right],
\end{split}
\end{equation}
where
\begin{equation}
\delta=\omega_p^2\gamma[(\omega\Gamma)^2+(\omega_{TO}^2-\omega^2)^2]+\omega^2\Gamma(\omega^2+\gamma^2)(\omega_{LO}^2-\omega_{TO}^2),
\end{equation}
where C is the Faust-Henry coefficient, $\omega_{LO}$ and $\omega_{TO}$ represent the LO and TO phonon frequencies of $A_1$ phonon mode, respectively. $\gamma$ and $\Gamma$ are the plasmon and phonon damping constants, respectively. $\omega_p$ is the plasma frequency given by following formula
\begin{equation}
\omega_p = \sqrt{\frac{4\pi ne^2}{\epsilon_{\infty}m^*}}
\end{equation}
where $n$ is the electron carrier density, $m^*$ is the effective mass of the electron and $\epsilon_{\infty}$ is the high frequency dielectric constant.
The dielectric function $\epsilon(\omega)$ in Eq.\eqref{I} is given by:
\begin{equation}
\epsilon(\omega)=\epsilon_{\infty}\left[1+\frac{\omega_{LO}^2-\omega_{TO}^2}{\omega_{TO}^2-\omega^2-\iota\omega\Gamma}-\frac{\omega_p^2}{\omega(\omega+\iota\gamma)}\right]
\end{equation}

\begin{figure}
\centering
\includegraphics[width=0.4\textwidth]{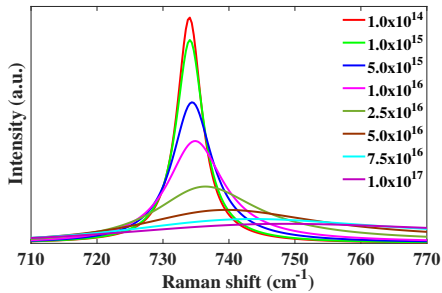}
\caption{Graphical representation of equation \eqref{I} with varying carrier concentration ($n$). The values of $n$ are in cm$^{-3}$ unit. Here the values of C, $\gamma$, $\Gamma$, $m^*$, $\epsilon_{\infty}$, $\omega_{TO}$ and $\omega_{LO}$  are taken as 0.48\cite{harima2002properties}, 8.5$\times 10^{13} \ s^{-1}$, 1.35$\times 10^{11} \ s^{-1} $, 0.18 $m_e$\cite{suzuki1995first}, 5.4\cite{melentev2016plasmon}, 531.8\cite{davydov1998phonon} and 734.0\cite{davydov1998phonon}, respectively.}
\label{matlab}
\end{figure}

The schematic representation of Eq.\ref{I} is shown in Fig.\ref{matlab}. The frequency of the LPP mode and its line width are dependent on carrier concentration and hence vary for samples with different carrier concentrations.  Increase in the peak  of Raman shift (frequency) and width of LPP mode, signifies increase in carrier density. For low carrier density  the character of LPP mode is phonon like due to reduced plasmon-phonon coupling. However,  with increasing carrier density the LPP peak shifts towards the higher frequency side, the intensity reduces and eventually the peak broadens due to increase in the coupling strength (see Figure \ref{matlab}).  Cheng \textit{et al.}\cite{cheng2009raman} have shown, in case of ZnO NRs, the LPP phonon peak broadening and shifting towards higher frequency with an increase in carrier concentration, consistent with our observations. Ding \textit{et al.}\cite{ding2012longitudinal} and Jeganathan \textit{et al.}\cite{jeganathan2009raman} have also reported similar behavior of LPP mode in Al-doped ZnO and Si-doped GaN, respectively.
\par
\begin{table*}
\centering
\caption{Peak position and line width of the LPP Raman modes obtained by deconvoluting the spectra of the samples. The $n_e$, $\gamma$ and $\Gamma$ are obtained by fitting the deconvoluted LPP mode with Eqn.\ref{I}.}
\setlength{\tabcolsep}{10pt}
\begin{tabular}{ c c c c c c c rrrr}
 \hline 
 \hline
 Sample & Mg:Ga & Peak & FWHM & $n_e$ & $\gamma$ & $\Gamma$ \\
 Name & & Position (cm$^{-1}$) & (cm$^{-1}$) & (cm$^{-3}$) & (THz) & (THz) \\
 \hline
 \hline
 A & 0.0000 & 737.3 & 30.9 & 1.5$\times$10$^{17}$ & 300 & 0.10 \\  
 B & 0.0082 & 738.8 & 17.0 & 4.5$\times10^{15}$ & 9 & 0.40 \\
 C & 0.0113 & 738.3 & 13.0 & 3.6$\times10^{15}$ & 1 & 0.40 \\
 D & 0.0162 & 739.3 & 9.7 & 5.0$\times10^{15}$ & 9 & 0.18  \\
 \hline
 \hline   
\end{tabular}
\label{raman_table}
\end{table*}

\begin{figure*}
\centering
\includegraphics[width=0.75\textwidth]{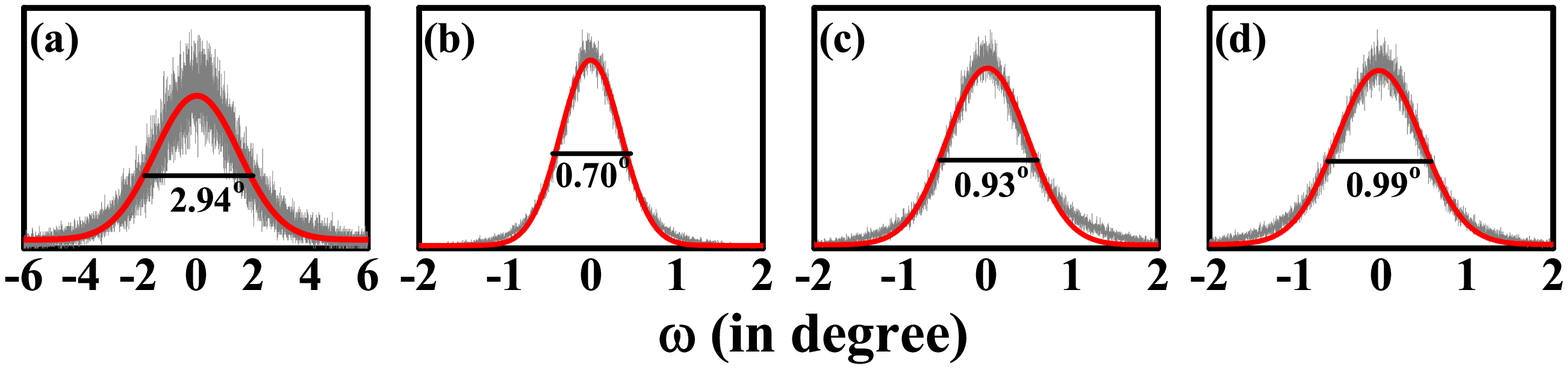}
\caption{(a), (b), (c) and (d) $\omega$-scans of samples A, B, C and D, respectively.}
\label{xrd}
\end{figure*}

\par
The peak position and FWHM of SO and LPP Raman modes is tabulated in Table \ref{raman_table}, where we note that the linewidth of LPP mode is strongly influenced by Mg concentration in the films. The deconvolated Raman spectra reveal that  the FWHMs of SO mode have a similar value of $\mathrm{\approx 20 \ cm^{-1}}$ for  all the sample under consideration (see Fig.\ref{LPP} (b)). We find that FWHM of LPP+ mode is significantly high ($\mathrm{\approx 30 cm^{-1}}$) in comparison to the Mg doped ones where the FWHM varied as 17.0, 13.0 and 9.7 $\mathrm{cm^{-1}}$ for sample B, C and D, respectively(see Fig.\ref{LPP} (b)). To make sure such changes are not due to any structural changes in the NRs samples, we carried out $\omega$-scan using HR-XRD studies whose width is a signature of the crystalline quality of the films. The FWHM of the (0002) planes as obtained from $\omega$-scan are noted as 2.94$\degree$, 0.70$\degree$, 0.93$\degree$ and 0.99$\degree$ for samples A, B, C and D respectively (see Fig.\ref{xrd}). Thus, there is no direct correlation between crystal quality and peak broadening of the LPP mode. Thus, the changes in the width of line shapes of Raman spectra becomes very significant  and can be used as a non-contact tool for the quantification of carrier concentration. We further fitted the deconvoluted LPP mode with Eqn.\ref{I} and the fitted plot is shown as solid lines with olive shade in Fig.\ref{LPP}(b). The obtained values of $n_e$, $\gamma$ and $\Gamma$ from the fitted curve are tabulated in Table \ref{raman_table}. Our estimate to the background carrier density for undoped GaN NRs sample (A) is 1.5$\times10^{17}$ cm$^{-3}$, which is in the same order of commonly observed carrier density in MBE grown GaN samples\cite{robins2016raman}. For Mg doped samples the background n-type carrier densities are estimated to the order of $10^{15}$  cm$^{-3}$ suggesting Mg incorporation has compensated the background carrier densities by two orders of magnitude. It is to note that for lightly doped sample B and C the background carrier density are 4.5$\times10^{15}$ and 3.6$\times10^{15}$  cm$^{-3}$ respectively whereas the background carrier density of D, which has relatively higher Mg concentration than B and C, is 5$\times10^{15}$  cm$^{-3}$ . At this stage we hypothesize that it could be the consequence of \textit{self-compensation} effect of the heavily Mg-doped GaN.
\par
The E$_2$(high) mode, being non-polar in nature, can be used as a measure of the inherent strain of the material. The FWHM of this mode reflects defect incorporation in the film, since strain gradient or phonon-defect scattering can lead to the broadening of this mode. We find a small change in the position of E$_2$ (high) peak (from 567.35 to 567.39 cm$^{-1}$) which suggests that a very small  macroscopic  strain is being  generated in the NRs with Mg incorporation. We also find that with the increase in Mg-flux, FWHM of E$_2$ (high) peak increases, which can be attributed to the generation of different kinds of point defects in the samples, depending on the concentration of Mg incorporation \cite{Kirste2013}. 
\begin{figure}
\centering
\includegraphics[width=0.4\textwidth]{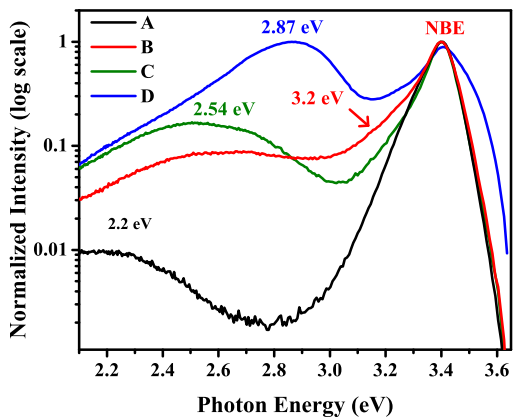}
\caption{Room temperature photoluminescence (PL) spectra of undoped and Mg doped GaN NRs.}
\label{PL}
\end{figure}

\par
The origin of the native $n$-type behavior of GaN is still ambiguous in the literature. Defects such as  N vacancies (V$_\mathrm{N}$) \cite{Boguslawski1995} and Oxygen substitution (O$_\mathrm{N}$) \cite{VanDeWalle2004} are reported as the major source of such auto-doping in GaN. Theoretical calculations based on \textit{first-principles} method further suggest that Oxygen substitution in the form of $\mathrm{V_{Ga}O_N}$ is energetically more favorable. \cite{neugebauer1996gallium}. Our previous work\cite{nayak2017vacancy} further suggests that defect formation energy of the V$_\mathrm{N}$ at the surface is much less than in the bulk and thus V$_\mathrm{N}$ is a major point defect in nanostructured GaN due to it's higher surface to volume ratio\cite{nayak2017vacancy}.
 Luminescence spectra recorded for sample A (undoped) shows a near band edge (NBE) peak at 3.4 eV and a broad YL at 2.20 eV (see Fig.\ref{PL}). 
However recent theoretical calculations  show that V$_\mathrm{N}$ and carbon related defects  may give rise to energy states in the bandgap responsible for yellow luminescence (YL) in PL spectra of n-GaN, whereas $\mathrm{V_{Ga}O_N}$ does not \cite{reshchikov2018thermal}. Thus, n-type and YL together suggest that V$_\mathrm{N}$ is the dominant defect type in this case. This speculation is further backed by the luminescence spectra obtained from sample B and C (moderately doped ones), wherein a green luminescence (GL) peak is observed at $\approx$ 2.54 eV along with NBE and a donor acceptor pair (DAP) transition at $\approx$ 3.2 eV (see Fig.\ref{PL}). The 3.2 eV luminescence peak is the consequence of the formation of the substitutional Mg in GaN (Mg$_\mathrm{Ga}$). Reshshikov \textit{et al.} \cite{reshchikov2014green} studied the defects in Mg doped GaN and suggested that most energetically favorable point defect is V$_\mathrm{N}$ and it gives rise to green luminescence, while defect complex $\mathrm{Mg_{Ga}V_N}$  results in red bands\cite{reshchikov2014green}. Our experimental PL data shows only GL band  and a DAP transition at 3.2 eV (sample B and C) thus confirming the formation of isolated V$_\mathrm{N}$ and $\mathrm{Mg_{Ga}}$ in Mg doped GaN NRs. Each isolated V$_\mathrm{N}$ results in one electron per vacancy, while $\mathrm{Mg_{Ga}}$ may results into a free hole. Thus, formation of individual V$_\mathrm{N}$ and $\mathrm{Mg_{Ga}}$, will not result in any change in carrier concentration.  SIMS measurements\cite{nayak2018enhanced} on these samples reveal Mg concentrations of 4.9$\times10^{19}$, 6.0$\times10^{19}$ and 2.9 $\times10^{20}$ atoms cm$^{-3}$ and the reduction in background carrier density ($n_e$) with the increase in Mg concentration suggesting that V$_\mathrm{N}$ concentration is less than 4.9$\times10^{19}$ cm$^{-3}$ in these samples. The compensation effect observed here is due to formation of isolated $\mathrm{Mg_{Ga}}$ which compensates the native n-type character caused by formation of isolated V$_\mathrm{N}$ in unintentionally n-doped GaN. 
\par 
In previous sections, we envisage that the higher value of background native $n$-type carrier density  in heavily Mg-doped sample D in comparison to B and C is due to the \textit{self-compensation} effect,  is further strengthened by the appearance of blue luminescence (BL) in PL spectra of sample D. In past we have identified that the BL is a signature of formation of the point defect complex Mg$_{Ga}$+Mg$_i$ and a n-type dopant to Mg-doped GaN\cite{nayak2017origin}. 
\par
In summary, we have grown Mg doped GaN NRs on Si (111) surface and studied the native charge compensation effect in the same using optical and X-ray photoelectron spectroscopies. Two order magnitude of native charge compensation was deduced upon Mg-doping with a concentration of 10$^{19}$-10$^{20}$ atoms cm$^{-3}$ by analysing the line shape profile of the longitudinal plasmon-phonon coupled Raman mode. Thus we propose that line shape of the longitudinal plasmon-phonon coupled Raman mode could be a contactless method in determining the charge carrier density and the dopant induced compensation in nanostructured materials. 

\bibliographystyle{apsrev}
\bibliography{bib}

\begin{thebibliography}{31}
\expandafter\ifx\csname natexlab\endcsname\relax\def\natexlab#1{#1}\fi
\expandafter\ifx\csname bibnamefont\endcsname\relax
  \def\bibnamefont#1{#1}\fi
\expandafter\ifx\csname bibfnamefont\endcsname\relax
  \def\bibfnamefont#1{#1}\fi
\expandafter\ifx\csname citenamefont\endcsname\relax
  \def\citenamefont#1{#1}\fi
\expandafter\ifx\csname url\endcsname\relax
  \def\url#1{\texttt{#1}}\fi
\expandafter\ifx\csname urlprefix\endcsname\relax\def\urlprefix{URL }\fi
\providecommand{\bibinfo}[2]{#2}
\providecommand{\eprint}[2][]{\url{#2}}

\bibitem[{\citenamefont{Ihn et~al.}(2006)\citenamefont{Ihn, Song, Kim, and
  Lee}}]{ihn2006gaas}
\bibinfo{author}{\bibfnamefont{S.-G.} \bibnamefont{Ihn}},
  \bibinfo{author}{\bibfnamefont{J.-I.} \bibnamefont{Song}},
  \bibinfo{author}{\bibfnamefont{Y.-H.} \bibnamefont{Kim}}, \bibnamefont{and}
  \bibinfo{author}{\bibfnamefont{J.~Y.} \bibnamefont{Lee}},
  \bibinfo{journal}{Appl. Phys. Lett.} \textbf{\bibinfo{volume}{89}},
  \bibinfo{pages}{053106} (\bibinfo{year}{2006}).

\bibitem[{\citenamefont{Li et~al.}(2008)\citenamefont{Li, Guimard, Rajesh, and
  Arakawa}}]{li2008growth}
\bibinfo{author}{\bibfnamefont{L.}~\bibnamefont{Li}},
  \bibinfo{author}{\bibfnamefont{D.}~\bibnamefont{Guimard}},
  \bibinfo{author}{\bibfnamefont{M.}~\bibnamefont{Rajesh}}, \bibnamefont{and}
  \bibinfo{author}{\bibfnamefont{Y.}~\bibnamefont{Arakawa}},
  \bibinfo{journal}{Appl. Phys. Lett.} \textbf{\bibinfo{volume}{92}},
  \bibinfo{pages}{263105} (\bibinfo{year}{2008}).

\bibitem[{\citenamefont{Cerutti et~al.}(2006)\citenamefont{Cerutti, Risti{\'c},
  Fern{\'a}ndez-Garrido, Calleja, Trampert, Ploog, Lazic, and
  Calleja}}]{cerutti2006wurtzite}
\bibinfo{author}{\bibfnamefont{L.}~\bibnamefont{Cerutti}},
  \bibinfo{author}{\bibfnamefont{J.}~\bibnamefont{Risti{\'c}}},
  \bibinfo{author}{\bibfnamefont{S.}~\bibnamefont{Fern{\'a}ndez-Garrido}},
  \bibinfo{author}{\bibfnamefont{E.}~\bibnamefont{Calleja}},
  \bibinfo{author}{\bibfnamefont{A.}~\bibnamefont{Trampert}},
  \bibinfo{author}{\bibfnamefont{K.}~\bibnamefont{Ploog}},
  \bibinfo{author}{\bibfnamefont{S.}~\bibnamefont{Lazic}}, \bibnamefont{and}
  \bibinfo{author}{\bibfnamefont{J.}~\bibnamefont{Calleja}},
  \bibinfo{journal}{Appl. Phys. Lett.} \textbf{\bibinfo{volume}{88}},
  \bibinfo{pages}{213114} (\bibinfo{year}{2006}).

\bibitem[{\citenamefont{M{\aa}rtensson
  et~al.}(2004)\citenamefont{M{\aa}rtensson, Svensson, Wacaser, Larsson,
  Seifert, Deppert, Gustafsson, Wallenberg, and
  Samuelson}}]{maartensson2004epitaxial}
\bibinfo{author}{\bibfnamefont{T.}~\bibnamefont{M{\aa}rtensson}},
  \bibinfo{author}{\bibfnamefont{C.~P.~T.} \bibnamefont{Svensson}},
  \bibinfo{author}{\bibfnamefont{B.~A.} \bibnamefont{Wacaser}},
  \bibinfo{author}{\bibfnamefont{M.~W.} \bibnamefont{Larsson}},
  \bibinfo{author}{\bibfnamefont{W.}~\bibnamefont{Seifert}},
  \bibinfo{author}{\bibfnamefont{K.}~\bibnamefont{Deppert}},
  \bibinfo{author}{\bibfnamefont{A.}~\bibnamefont{Gustafsson}},
  \bibinfo{author}{\bibfnamefont{L.~R.} \bibnamefont{Wallenberg}},
  \bibnamefont{and}
  \bibinfo{author}{\bibfnamefont{L.}~\bibnamefont{Samuelson}},
  \bibinfo{journal}{Nano Lett.} \textbf{\bibinfo{volume}{4}},
  \bibinfo{pages}{1987} (\bibinfo{year}{2004}).

\bibitem[{\citenamefont{Glas}(2006)}]{glas2006critical}
\bibinfo{author}{\bibfnamefont{F.}~\bibnamefont{Glas}}, \bibinfo{journal}{Phys.
  Rev. B} \textbf{\bibinfo{volume}{74}}, \bibinfo{pages}{121302}
  (\bibinfo{year}{2006}).

\bibitem[{\citenamefont{Verheijen et~al.}(2006)\citenamefont{Verheijen, Immink,
  de~Smet, Borgstr{\"o}m, and Bakkers}}]{verheijen2006growth}
\bibinfo{author}{\bibfnamefont{M.~A.} \bibnamefont{Verheijen}},
  \bibinfo{author}{\bibfnamefont{G.}~\bibnamefont{Immink}},
  \bibinfo{author}{\bibfnamefont{T.}~\bibnamefont{de~Smet}},
  \bibinfo{author}{\bibfnamefont{M.~T.} \bibnamefont{Borgstr{\"o}m}},
  \bibnamefont{and} \bibinfo{author}{\bibfnamefont{E.~P.}
  \bibnamefont{Bakkers}}, \bibinfo{journal}{J. Am. Chem. Soc.}
  \textbf{\bibinfo{volume}{128}}, \bibinfo{pages}{1353} (\bibinfo{year}{2006}).

\bibitem[{\citenamefont{Lin et~al.}(2010)\citenamefont{Lin, Lu, Chen, Lee, ,
  and Gwo}}]{lin2010ingan}
\bibinfo{author}{\bibfnamefont{H.-W.} \bibnamefont{Lin}},
  \bibinfo{author}{\bibfnamefont{Y.-J.} \bibnamefont{Lu}},
  \bibinfo{author}{\bibfnamefont{H.-Y.} \bibnamefont{Chen}},
  \bibinfo{author}{\bibfnamefont{H.-M.} \bibnamefont{Lee}}, , \bibnamefont{and}
  \bibinfo{author}{\bibfnamefont{S.}~\bibnamefont{Gwo}},
  \bibinfo{journal}{Appl. Phys. Lett.} \textbf{\bibinfo{volume}{97}},
  \bibinfo{pages}{073101} (\bibinfo{year}{2010}).

\bibitem[{\citenamefont{Bengoechea-Encabo
  et~al.}(2014)\citenamefont{Bengoechea-Encabo, Albert, Lopez-Romero, Lefebvre,
  Barbagini, Torres-Pardo, Gonz{\'a}lez-Calbet, Sanchez-Garcia, and
  Calleja}}]{bengoechea2014light}
\bibinfo{author}{\bibfnamefont{A.}~\bibnamefont{Bengoechea-Encabo}},
  \bibinfo{author}{\bibfnamefont{S.}~\bibnamefont{Albert}},
  \bibinfo{author}{\bibfnamefont{D.}~\bibnamefont{Lopez-Romero}},
  \bibinfo{author}{\bibfnamefont{P.}~\bibnamefont{Lefebvre}},
  \bibinfo{author}{\bibfnamefont{F.}~\bibnamefont{Barbagini}},
  \bibinfo{author}{\bibfnamefont{A.}~\bibnamefont{Torres-Pardo}},
  \bibinfo{author}{\bibfnamefont{J.~M.} \bibnamefont{Gonz{\'a}lez-Calbet}},
  \bibinfo{author}{\bibfnamefont{M.~A.} \bibnamefont{Sanchez-Garcia}},
  \bibnamefont{and} \bibinfo{author}{\bibfnamefont{E.}~\bibnamefont{Calleja}},
  \bibinfo{journal}{Nanotechnology} \textbf{\bibinfo{volume}{25}},
  \bibinfo{pages}{435203} (\bibinfo{year}{2014}).

\bibitem[{\citenamefont{Wang et~al.}(2008)\citenamefont{Wang, Chen, Chen,
  Cheng, Ke, Hsieh, Wu, Peng, and Huang}}]{wang2008gan}
\bibinfo{author}{\bibfnamefont{C.-Y.} \bibnamefont{Wang}},
  \bibinfo{author}{\bibfnamefont{L.-Y.} \bibnamefont{Chen}},
  \bibinfo{author}{\bibfnamefont{C.-P.} \bibnamefont{Chen}},
  \bibinfo{author}{\bibfnamefont{Y.-W.} \bibnamefont{Cheng}},
  \bibinfo{author}{\bibfnamefont{M.-Y.} \bibnamefont{Ke}},
  \bibinfo{author}{\bibfnamefont{M.-Y.} \bibnamefont{Hsieh}},
  \bibinfo{author}{\bibfnamefont{H.-M.} \bibnamefont{Wu}},
  \bibinfo{author}{\bibfnamefont{L.-H.} \bibnamefont{Peng}}, \bibnamefont{and}
  \bibinfo{author}{\bibfnamefont{J.}~\bibnamefont{Huang}},
  \bibinfo{journal}{Opt. Express} \textbf{\bibinfo{volume}{16}},
  \bibinfo{pages}{10549} (\bibinfo{year}{2008}).

\bibitem[{\citenamefont{Nguyen et~al.}(2011)\citenamefont{Nguyen, Zhang, Cui,
  Han, Fathololoumi, Couillard, Botton, and Mi}}]{nguyen2011p}
\bibinfo{author}{\bibfnamefont{H.~P.~T.} \bibnamefont{Nguyen}},
  \bibinfo{author}{\bibfnamefont{S.}~\bibnamefont{Zhang}},
  \bibinfo{author}{\bibfnamefont{K.}~\bibnamefont{Cui}},
  \bibinfo{author}{\bibfnamefont{X.}~\bibnamefont{Han}},
  \bibinfo{author}{\bibfnamefont{S.}~\bibnamefont{Fathololoumi}},
  \bibinfo{author}{\bibfnamefont{M.}~\bibnamefont{Couillard}},
  \bibinfo{author}{\bibfnamefont{G.}~\bibnamefont{Botton}}, \bibnamefont{and}
  \bibinfo{author}{\bibfnamefont{Z.}~\bibnamefont{Mi}}, \bibinfo{journal}{Nano
  Lett.} \textbf{\bibinfo{volume}{11}}, \bibinfo{pages}{1919}
  (\bibinfo{year}{2011}).

\bibitem[{\citenamefont{Wang et~al.}(2014)\citenamefont{Wang, Liu, Kibria,
  Zhao, Nguyen, Li, Mi, Gonzalez, and Andrews}}]{wang2014p}
\bibinfo{author}{\bibfnamefont{Q.}~\bibnamefont{Wang}},
  \bibinfo{author}{\bibfnamefont{X.}~\bibnamefont{Liu}},
  \bibinfo{author}{\bibfnamefont{M.}~\bibnamefont{Kibria}},
  \bibinfo{author}{\bibfnamefont{S.}~\bibnamefont{Zhao}},
  \bibinfo{author}{\bibfnamefont{H.}~\bibnamefont{Nguyen}},
  \bibinfo{author}{\bibfnamefont{K.}~\bibnamefont{Li}},
  \bibinfo{author}{\bibfnamefont{Z.}~\bibnamefont{Mi}},
  \bibinfo{author}{\bibfnamefont{T.}~\bibnamefont{Gonzalez}}, \bibnamefont{and}
  \bibinfo{author}{\bibfnamefont{M.}~\bibnamefont{Andrews}},
  \bibinfo{journal}{Nanoscale} \textbf{\bibinfo{volume}{6}},
  \bibinfo{pages}{9970} (\bibinfo{year}{2014}).

\bibitem[{\citenamefont{Nayak et~al.}(2018)\citenamefont{Nayak, Kumar, Pandey,
  Nagaraja, Gupta, and Shivaprasad}}]{nayak2018enhanced}
\bibinfo{author}{\bibfnamefont{S.}~\bibnamefont{Nayak}},
  \bibinfo{author}{\bibfnamefont{R.}~\bibnamefont{Kumar}},
  \bibinfo{author}{\bibfnamefont{N.}~\bibnamefont{Pandey}},
  \bibinfo{author}{\bibfnamefont{K.}~\bibnamefont{Nagaraja}},
  \bibinfo{author}{\bibfnamefont{M.}~\bibnamefont{Gupta}}, \bibnamefont{and}
  \bibinfo{author}{\bibfnamefont{S.}~\bibnamefont{Shivaprasad}},
  \bibinfo{journal}{J. Appl. Phys.} \textbf{\bibinfo{volume}{123}},
  \bibinfo{pages}{135303} (\bibinfo{year}{2018}).

\bibitem[{\citenamefont{Kaschner et~al.}(1999)\citenamefont{Kaschner, Siegle,
  Kaczmarczyk, Stra{\ss}burg, Hoffmann, Thomsen, Birkle, Einfeldt, and
  Hommel}}]{Kaschner1999}
\bibinfo{author}{\bibfnamefont{A.}~\bibnamefont{Kaschner}},
  \bibinfo{author}{\bibfnamefont{H.}~\bibnamefont{Siegle}},
  \bibinfo{author}{\bibfnamefont{G.}~\bibnamefont{Kaczmarczyk}},
  \bibinfo{author}{\bibfnamefont{M.}~\bibnamefont{Stra{\ss}burg}},
  \bibinfo{author}{\bibfnamefont{A.}~\bibnamefont{Hoffmann}},
  \bibinfo{author}{\bibfnamefont{C.}~\bibnamefont{Thomsen}},
  \bibinfo{author}{\bibfnamefont{U.}~\bibnamefont{Birkle}},
  \bibinfo{author}{\bibfnamefont{S.}~\bibnamefont{Einfeldt}}, \bibnamefont{and}
  \bibinfo{author}{\bibfnamefont{D.}~\bibnamefont{Hommel}},
  \bibinfo{journal}{Appl. Phys. Lett.} \textbf{\bibinfo{volume}{74}},
  \bibinfo{pages}{3281} (\bibinfo{year}{1999}).

\bibitem[{\citenamefont{Davydov et~al.}(1998)\citenamefont{Davydov, Kitaev,
  Goncharuk, Smirnov, Graul, Semchinova, Uffmann, Smirnov, Mirgorodsky, and
  Evarestov}}]{davydov1998phonon}
\bibinfo{author}{\bibfnamefont{V.~Y.} \bibnamefont{Davydov}},
  \bibinfo{author}{\bibfnamefont{Y.~E.} \bibnamefont{Kitaev}},
  \bibinfo{author}{\bibfnamefont{I.}~\bibnamefont{Goncharuk}},
  \bibinfo{author}{\bibfnamefont{A.}~\bibnamefont{Smirnov}},
  \bibinfo{author}{\bibfnamefont{J.}~\bibnamefont{Graul}},
  \bibinfo{author}{\bibfnamefont{O.}~\bibnamefont{Semchinova}},
  \bibinfo{author}{\bibfnamefont{D.}~\bibnamefont{Uffmann}},
  \bibinfo{author}{\bibfnamefont{M.}~\bibnamefont{Smirnov}},
  \bibinfo{author}{\bibfnamefont{A.}~\bibnamefont{Mirgorodsky}},
  \bibnamefont{and}
  \bibinfo{author}{\bibfnamefont{R.}~\bibnamefont{Evarestov}},
  \bibinfo{journal}{Phys. Rev. B} \textbf{\bibinfo{volume}{58}},
  \bibinfo{pages}{12899} (\bibinfo{year}{1998}).

\bibitem[{\citenamefont{Azuhata et~al.}(1995)\citenamefont{Azuhata, Sota,
  Suzuki, and Nakamura}}]{azuhata1995polarized}
\bibinfo{author}{\bibfnamefont{T.}~\bibnamefont{Azuhata}},
  \bibinfo{author}{\bibfnamefont{T.}~\bibnamefont{Sota}},
  \bibinfo{author}{\bibfnamefont{K.}~\bibnamefont{Suzuki}}, \bibnamefont{and}
  \bibinfo{author}{\bibfnamefont{S.}~\bibnamefont{Nakamura}},
  \bibinfo{journal}{J. Phys. Condens. Matter} \textbf{\bibinfo{volume}{7}},
  \bibinfo{pages}{L129} (\bibinfo{year}{1995}).

\bibitem[{\citenamefont{Siegle et~al.}(1997)\citenamefont{Siegle, Kaczmarczyk,
  Filippidis, Litvinchuk, Hoffmann, and Thomsen}}]{siegle1997zone}
\bibinfo{author}{\bibfnamefont{H.}~\bibnamefont{Siegle}},
  \bibinfo{author}{\bibfnamefont{G.}~\bibnamefont{Kaczmarczyk}},
  \bibinfo{author}{\bibfnamefont{L.}~\bibnamefont{Filippidis}},
  \bibinfo{author}{\bibfnamefont{A.}~\bibnamefont{Litvinchuk}},
  \bibinfo{author}{\bibfnamefont{A.}~\bibnamefont{Hoffmann}}, \bibnamefont{and}
  \bibinfo{author}{\bibfnamefont{C.}~\bibnamefont{Thomsen}},
  \bibinfo{journal}{Phys. Rev. B} \textbf{\bibinfo{volume}{55}},
  \bibinfo{pages}{7000} (\bibinfo{year}{1997}).

\bibitem[{\citenamefont{Robins et~al.}(2016)\citenamefont{Robins, Horneber,
  Sanford, Bertness, Brubaker, and Schlager}}]{robins2016raman}
\bibinfo{author}{\bibfnamefont{L.~H.} \bibnamefont{Robins}},
  \bibinfo{author}{\bibfnamefont{E.}~\bibnamefont{Horneber}},
  \bibinfo{author}{\bibfnamefont{N.~A.} \bibnamefont{Sanford}},
  \bibinfo{author}{\bibfnamefont{K.~A.} \bibnamefont{Bertness}},
  \bibinfo{author}{\bibfnamefont{M.}~\bibnamefont{Brubaker}}, \bibnamefont{and}
  \bibinfo{author}{\bibfnamefont{J.}~\bibnamefont{Schlager}},
  \bibinfo{journal}{J. Appl. Phys.} \textbf{\bibinfo{volume}{120}},
  \bibinfo{pages}{124313} (\bibinfo{year}{2016}).

\bibitem[{\citenamefont{Cheng et~al.}(2009)\citenamefont{Cheng, Tzeng, Xu,
  Alur, Wang, Park, Wu, Shannon, Kim, and Wang}}]{cheng2009raman}
\bibinfo{author}{\bibfnamefont{A.-J.} \bibnamefont{Cheng}},
  \bibinfo{author}{\bibfnamefont{Y.}~\bibnamefont{Tzeng}},
  \bibinfo{author}{\bibfnamefont{H.}~\bibnamefont{Xu}},
  \bibinfo{author}{\bibfnamefont{S.}~\bibnamefont{Alur}},
  \bibinfo{author}{\bibfnamefont{Y.}~\bibnamefont{Wang}},
  \bibinfo{author}{\bibfnamefont{M.}~\bibnamefont{Park}},
  \bibinfo{author}{\bibfnamefont{T.-h.} \bibnamefont{Wu}},
  \bibinfo{author}{\bibfnamefont{C.}~\bibnamefont{Shannon}},
  \bibinfo{author}{\bibfnamefont{D.-J.} \bibnamefont{Kim}}, \bibnamefont{and}
  \bibinfo{author}{\bibfnamefont{D.}~\bibnamefont{Wang}}, \bibinfo{journal}{J.
  Appl. Phys.} \textbf{\bibinfo{volume}{105}}, \bibinfo{pages}{073104}
  (\bibinfo{year}{2009}).

\bibitem[{\citenamefont{Ding et~al.}(2012)\citenamefont{Ding, Hu, Lin, Huang,
  and Huang}}]{ding2012longitudinal}
\bibinfo{author}{\bibfnamefont{K.}~\bibnamefont{Ding}},
  \bibinfo{author}{\bibfnamefont{Q.}~\bibnamefont{Hu}},
  \bibinfo{author}{\bibfnamefont{W.}~\bibnamefont{Lin}},
  \bibinfo{author}{\bibfnamefont{J.}~\bibnamefont{Huang}}, \bibnamefont{and}
  \bibinfo{author}{\bibfnamefont{F.}~\bibnamefont{Huang}},
  \bibinfo{journal}{Appl. Phys. Lett.} \textbf{\bibinfo{volume}{101}},
  \bibinfo{pages}{031908} (\bibinfo{year}{2012}).

\bibitem[{\citenamefont{Jeganathan et~al.}(2009)\citenamefont{Jeganathan,
  Debnath, Meijers, Stoica, Calarco, Gr{\"u}tzmacher, and
  L{\"u}th}}]{jeganathan2009raman}
\bibinfo{author}{\bibfnamefont{K.}~\bibnamefont{Jeganathan}},
  \bibinfo{author}{\bibfnamefont{R.}~\bibnamefont{Debnath}},
  \bibinfo{author}{\bibfnamefont{R.}~\bibnamefont{Meijers}},
  \bibinfo{author}{\bibfnamefont{T.}~\bibnamefont{Stoica}},
  \bibinfo{author}{\bibfnamefont{R.}~\bibnamefont{Calarco}},
  \bibinfo{author}{\bibfnamefont{D.}~\bibnamefont{Gr{\"u}tzmacher}},
  \bibnamefont{and} \bibinfo{author}{\bibfnamefont{H.}~\bibnamefont{L{\"u}th}},
  \bibinfo{journal}{J. Appl. Phys.} \textbf{\bibinfo{volume}{105}},
  \bibinfo{pages}{123707} (\bibinfo{year}{2009}).

\bibitem[{\citenamefont{Harima}(2002)}]{harima2002properties}
\bibinfo{author}{\bibfnamefont{H.}~\bibnamefont{Harima}}, \bibinfo{journal}{J.
  Phys. Condens. Matter} \textbf{\bibinfo{volume}{14}}, \bibinfo{pages}{R967}
  (\bibinfo{year}{2002}).

\bibitem[{\citenamefont{Suzuki et~al.}(1995)\citenamefont{Suzuki, Uenoyama, and
  Yanase}}]{suzuki1995first}
\bibinfo{author}{\bibfnamefont{M.}~\bibnamefont{Suzuki}},
  \bibinfo{author}{\bibfnamefont{T.}~\bibnamefont{Uenoyama}}, \bibnamefont{and}
  \bibinfo{author}{\bibfnamefont{A.}~\bibnamefont{Yanase}},
  \bibinfo{journal}{Phys. Rev. B} \textbf{\bibinfo{volume}{52}},
  \bibinfo{pages}{8132} (\bibinfo{year}{1995}).

\bibitem[{\citenamefont{Melentev et~al.}(2016)\citenamefont{Melentev,
  Yaichnikov, Shalygin, Vinnichenko, Vorobjev, Firsov, Riuttanen, and
  Suihkonen}}]{melentev2016plasmon}
\bibinfo{author}{\bibfnamefont{G.}~\bibnamefont{Melentev}},
  \bibinfo{author}{\bibfnamefont{D.~Y.} \bibnamefont{Yaichnikov}},
  \bibinfo{author}{\bibfnamefont{V.}~\bibnamefont{Shalygin}},
  \bibinfo{author}{\bibfnamefont{M.~Y.} \bibnamefont{Vinnichenko}},
  \bibinfo{author}{\bibfnamefont{L.}~\bibnamefont{Vorobjev}},
  \bibinfo{author}{\bibfnamefont{D.}~\bibnamefont{Firsov}},
  \bibinfo{author}{\bibfnamefont{L.}~\bibnamefont{Riuttanen}},
  \bibnamefont{and}
  \bibinfo{author}{\bibfnamefont{S.}~\bibnamefont{Suihkonen}}, in
  \emph{\bibinfo{booktitle}{J. Phys. Conf. Ser}} (\bibinfo{organization}{IOP
  Publishing}, \bibinfo{year}{2016}), vol. \bibinfo{volume}{690}, p.
  \bibinfo{pages}{012005}.

\bibitem[{\citenamefont{Kirste et~al.}(2013)\citenamefont{Kirste, Hoffmann,
  Tweedie, Bryan, Callsen, Kure, Nenstiel, Wagner, Collazo, Hoffmann
  et~al.}}]{Kirste2013}
\bibinfo{author}{\bibfnamefont{R.}~\bibnamefont{Kirste}},
  \bibinfo{author}{\bibfnamefont{M.~P.} \bibnamefont{Hoffmann}},
  \bibinfo{author}{\bibfnamefont{J.}~\bibnamefont{Tweedie}},
  \bibinfo{author}{\bibfnamefont{Z.}~\bibnamefont{Bryan}},
  \bibinfo{author}{\bibfnamefont{G.}~\bibnamefont{Callsen}},
  \bibinfo{author}{\bibfnamefont{T.}~\bibnamefont{Kure}},
  \bibinfo{author}{\bibfnamefont{C.}~\bibnamefont{Nenstiel}},
  \bibinfo{author}{\bibfnamefont{M.~R.} \bibnamefont{Wagner}},
  \bibinfo{author}{\bibfnamefont{R.}~\bibnamefont{Collazo}},
  \bibinfo{author}{\bibfnamefont{A.}~\bibnamefont{Hoffmann}},
  \bibnamefont{et~al.}, \bibinfo{journal}{J. Appl. Phys.}
  \textbf{\bibinfo{volume}{113}}, \bibinfo{pages}{103504}
  (\bibinfo{year}{2013}).

\bibitem[{\citenamefont{Boguslawski et~al.}(1995)\citenamefont{Boguslawski,
  Briggs, and Bernholc}}]{Boguslawski1995}
\bibinfo{author}{\bibfnamefont{P.}~\bibnamefont{Boguslawski}},
  \bibinfo{author}{\bibfnamefont{E.~L.} \bibnamefont{Briggs}},
  \bibnamefont{and} \bibinfo{author}{\bibfnamefont{J.}~\bibnamefont{Bernholc}},
  \bibinfo{journal}{Phys. Rev. B} \textbf{\bibinfo{volume}{51}},
  \bibinfo{pages}{17255} (\bibinfo{year}{1995}).

\bibitem[{\citenamefont{{Van De Walle} and Neugebauer}(2004)}]{VanDeWalle2004}
\bibinfo{author}{\bibfnamefont{C.~G.} \bibnamefont{{Van De Walle}}}
  \bibnamefont{and}
  \bibinfo{author}{\bibfnamefont{J.}~\bibnamefont{Neugebauer}},
  \bibinfo{journal}{J. Appl. Phys.} \textbf{\bibinfo{volume}{95}},
  \bibinfo{pages}{3851} (\bibinfo{year}{2004}).

\bibitem[{\citenamefont{Neugebauer and Van~de
  Walle}(1996)}]{neugebauer1996gallium}
\bibinfo{author}{\bibfnamefont{J.}~\bibnamefont{Neugebauer}} \bibnamefont{and}
  \bibinfo{author}{\bibfnamefont{C.~G.} \bibnamefont{Van~de Walle}},
  \bibinfo{journal}{Appl. Phys. Lett.} \textbf{\bibinfo{volume}{69}},
  \bibinfo{pages}{503} (\bibinfo{year}{1996}).

\bibitem[{\citenamefont{Reshchikov et~al.}(2018)\citenamefont{Reshchikov,
  Albarakati, Monavarian, Avrutin, and Morkoc}}]{reshchikov2018thermal}
\bibinfo{author}{\bibfnamefont{M.}~\bibnamefont{Reshchikov}},
  \bibinfo{author}{\bibfnamefont{N.}~\bibnamefont{Albarakati}},
  \bibinfo{author}{\bibfnamefont{M.}~\bibnamefont{Monavarian}},
  \bibinfo{author}{\bibfnamefont{V.}~\bibnamefont{Avrutin}}, \bibnamefont{and}
  \bibinfo{author}{\bibfnamefont{H.}~\bibnamefont{Morkoc}},
  \bibinfo{journal}{J. Appl. Phys.} \textbf{\bibinfo{volume}{123}},
  \bibinfo{pages}{161520} (\bibinfo{year}{2018}).

\bibitem[{\citenamefont{Nayak et~al.}(2017{\natexlab{a}})\citenamefont{Nayak,
  Naik, Jain, Waghmare, and Shivaprasad}}]{nayak2017vacancy}
\bibinfo{author}{\bibfnamefont{S.}~\bibnamefont{Nayak}},
  \bibinfo{author}{\bibfnamefont{M.~H.} \bibnamefont{Naik}},
  \bibinfo{author}{\bibfnamefont{M.}~\bibnamefont{Jain}},
  \bibinfo{author}{\bibfnamefont{U.}~\bibnamefont{Waghmare}}, \bibnamefont{and}
  \bibinfo{author}{\bibfnamefont{S.}~\bibnamefont{Shivaprasad}},
  \bibinfo{journal}{arXiv preprint arXiv:1710.05670}
  (\bibinfo{year}{2017}{\natexlab{a}}).

\bibitem[{\citenamefont{Reshchikov et~al.}(2014)\citenamefont{Reshchikov,
  Demchenko, McNamara, Fern{\'a}ndez-Garrido, and
  Calarco}}]{reshchikov2014green}
\bibinfo{author}{\bibfnamefont{M.~A.} \bibnamefont{Reshchikov}},
  \bibinfo{author}{\bibfnamefont{D.}~\bibnamefont{Demchenko}},
  \bibinfo{author}{\bibfnamefont{J.}~\bibnamefont{McNamara}},
  \bibinfo{author}{\bibfnamefont{S.}~\bibnamefont{Fern{\'a}ndez-Garrido}},
  \bibnamefont{and} \bibinfo{author}{\bibfnamefont{R.}~\bibnamefont{Calarco}},
  \bibinfo{journal}{Phys. Rev. B} \textbf{\bibinfo{volume}{90}},
  \bibinfo{pages}{035207} (\bibinfo{year}{2014}).

\bibitem[{\citenamefont{Nayak et~al.}(2017{\natexlab{b}})\citenamefont{Nayak,
  Gupta, Waghmare, and Shivaprasad}}]{nayak2017origin}
\bibinfo{author}{\bibfnamefont{S.}~\bibnamefont{Nayak}},
  \bibinfo{author}{\bibfnamefont{M.}~\bibnamefont{Gupta}},
  \bibinfo{author}{\bibfnamefont{U.~V.} \bibnamefont{Waghmare}},
  \bibnamefont{and}
  \bibinfo{author}{\bibfnamefont{S.}~\bibnamefont{Shivaprasad}},
  \bibinfo{journal}{arXiv preprint arXiv:1708.04036}
  (\bibinfo{year}{2017}{\natexlab{b}}).

\end{thebibliography}
\end{document}